\documentclass[acmsmall, screen]{acmart}

\usepackage{syntax}
\usepackage{booktabs} % For formal tables
\usepackage{mathtools}
\usepackage{forest}
\usepackage{subcaption}
\usepackage[ruled]{algorithm2e} % For algorithms
\usepackage{svg} % For svg images
\usepackage{fancyvrb}
\usepackage{cleveref}

\usepackage{tikz}

\SetAlFnt{\small}
\SetAlCapFnt{\small}
\SetAlCapNameFnt{\small}
\SetAlCapHSkip{0pt}
\IncMargin{-\parindent}

\usepackage{xcolor}

\setcopyright{cc}
\setcctype{by}
\acmDOI{10.1145/3729322}
\acmYear{2025}
\acmJournal{PACMPL}
\acmVolume{9}
\acmNumber{PLDI}
\acmArticle{219}
\acmMonth{6}
\received{2024-11-15}
\received[accepted]{2025-03-06}

\begin{document}

\title{Spineless Traversal for Layout Invalidation}

\author{Marisa Kirisame}
\orcid{0000-0002-3418-4835}
\affiliation{%
  \institution{University of Utah}
  \city{Salt Lake City}
  \country{USA}
}
\email{marisa@cs.utah.edu}

\author{Tiezhi Wang}
\orcid{0009-0003-7002-6011}
\affiliation{%
  \institution{Tongji University}
  \city{Shanghai}
  \country{China}
}
\email{2152591@tongji.edu.cn}

\author{Pavel Panchekha}
\orcid{0000-0003-2621-3592}
\affiliation{%
  \institution{University of Utah}
  \city{Salt Lake City}
  \country{USA}
}
\email{pavpan@cs.utah.edu}

\begin{abstract}
Latency is a major concern for web rendering engines
  like those in Chrome, Safari, and Firefox.
These engines reduce latency by using
  an \emph{incremental layout algorithm}
  to redraw the page when
  the user interacts with it.
In such an algorithm,
  elements that change frame-to-frame are marked dirty, and
  only those elements are processed
  to draw the next frame,
  dramatically reducing latency.
However, the standard incremental layout algorithm
  must search the page for dirty elements,
  accessing auxiliary elements in the process.
These auxiliary elements
  add cache misses and stalled cycles,
  and are responsible for a sizable fraction
  of all layout latency.

We introduce a new, faster incremental layout algorithm
  called Spineless Traversal.
Spineless Traversal
  uses a cache-friendlier priority queue algorithm
  that avoids accessing auxiliary nodes
  and thus reduces cache traffic and stalls.
This leads to dramatic speedups
  on the most latency-critical interactions
  such as hovering, typing, and animation.
Moreover, thanks to numerous low-level optimizations,
  Spineless Traversal is competitive
  across the whole spectrum of incremental layout workloads.
Spineless Traversal is faster than the standard approach
  on \PctFaster of \NumFrames~benchmarks,
  with a mean speedup of \MeanSpeedup
  concentrated in the most latency-critical interactions.
\end{abstract}

\newcommand{\DBPQoverheadCount}{2216}
\newcommand{\DBPQoverheadpctslowdown}{17.0\%\xspace}
\newcommand{\DBPQoverheadpctspeedup}{83.0\%\xspace}
\newcommand{\DBPQoverhead}{3.10}
\newcommand{\DBPQeval}{1.00}
\newcommand{\DBPQtotalCount}{2216}
\newcommand{\DBPQtotalpctslowdown}{17.0\%\xspace}
\newcommand{\DBPQtotalpctspeedup}{83.0\%\xspace}
\newcommand{\DBPQtotal}{\ensuremath{1.80\times}\xspace}
\newcommand{\DBPQsmalloverheadCount}{1454}
\newcommand{\DBPQsmalloverheadpctslowdown}{12.0\%}
\newcommand{\DBPQsmalloverheadpctspeedup}{88.0\%}
\newcommand{\DBPQsmalloverhead}{4.81}
\newcommand{\DBPQsmalleval}{1.00}
\newcommand{\DBPQsmalltotalCount}{1454}
\newcommand{\DBPQsmalltotalpctslowdown}{11.3\%}
\newcommand{\DBPQsmalltotalpctspeedup}{88.7\%}
\newcommand{\DBPQsmalltotal}{\ensuremath{2.22\times}\xspace}
\newcommand{\DBPQlargeoverhead}{1.34}
\newcommand{\DBPQlargeeval}{1.01}
\newcommand{\DBPQlargetotal}{1.20}
\newcommand{\TotalDiffCount}{2216\xspace}
\newcommand{\TotalTraceCount}{50\xspace}

\newcommand{\NumWebsites}{\TotalTraceCount}
\newcommand{\NumFrames}{\TotalDiffCount}
\newcommand{\MeanSpeedup}{\DBPQtotal}
\newcommand{\MeanSpeedupSmall}{\DBPQsmalltotal}
\newcommand{\PctSlower}{\DBPQtotalpctslowdown}
\newcommand{\PctFaster}{\DBPQtotalpctspeedup}
\newcommand{\PctSmall}{65.6\%\xspace}
\newcommand{\PctSlowerSmall}{10\%\xspace}

%
% Generate CCS codes using http://dl.acm.org/ccs.cfm and paste below
%
\begin{CCSXML}
<ccs2012>
   <concept>
       <concept_id>10011007.10011006.10011041.10011047</concept_id>
       <concept_desc>Software and its engineering~Source code generation</concept_desc>
       <concept_significance>500</concept_significance>
       </concept>
   <concept>
       <concept_id>10011007.10011006.10011041.10011046</concept_id>
       <concept_desc>Software and its engineering~Translator writing systems and compiler generators</concept_desc>
       <concept_significance>500</concept_significance>
       </concept>
   <concept>
       <concept_id>10011007.10011006.10011050.10011017</concept_id>
       <concept_desc>Software and its engineering~Domain specific languages</concept_desc>
       <concept_significance>500</concept_significance>
       </concept>
   <concept>
       <concept_id>10003752.10003809</concept_id>
       <concept_desc>Theory of computation~Design and analysis of algorithms</concept_desc>
       <concept_significance>500</concept_significance>
       </concept>
 </ccs2012>
\end{CCSXML}

\ccsdesc[500]{Software and its engineering~Source code generation}
\ccsdesc[500]{Software and its engineering~Translator writing systems and compiler generators}
\ccsdesc[500]{Software and its engineering~Domain specific languages}
\ccsdesc[500]{Theory of computation~Design and analysis of algorithms}

% We no longer use \terms command
%\terms{Design, Algorithms, Performance}

\keywords{Web Browsers, Layout, Incremental Computing, Order Maintenance, Latency}

\maketitle

\renewcommand{\shortauthors}{Marisa Kirisame, Tiezhi Wang, and Pavel Panchekha}
\section{Introduction}

Latency is a major concern for modern web rendering engines.
A rendering engine, such as that in Chrome, Firefox, or Safari,
  must redraw pages 60 times per second
  to guarantee smooth animations, fluid interactions,
  and responsive web applications.
When this frame rate cannot be met,
  the user experiences lag and may be forced to use another web application, browser, or device.
Modern 120\,Hz displays demand even lower latency.

Layout is a key driver of web rendering latency.
Layout means calculating the size and position
  of each element on the web page,
  after which the page can be rendered as pixels on the screen.
Every time the user interacts with the web page
  by hovering over an element,
  receiving updated data,
  or even observing an animation,
  the web page changes
   and must be re-laid-out.
Since layout is only one part of the larger rendering pipeline,
  this re-layout must be completed in a millisecond or less
  in order to meet the 60 frame-per-second goal.
On such a tight budget, every cycle counts!

\paragraph{Incrementalization}
The key optimization that makes this possible
  is \emph{incrementalization}.
When an element on the page changes,
  the browser \emph{marks} it dirty.
When the page is re-laid-out,
  the rendering engine traverses the page
  to find and re-lay-out only the dirty elements.
That might mark additional elements dirty,
  due to dependencies between elements,
  but typically---in, say, animations---%
  only a few nodes are ultimately re-laid-out.
In these cases, searching the tree
  for dirty elements is the bottleneck,
  especially since every element access
  is likely to incur a cache miss.

Browsers use the ``Double Dirty Bit'' algorithm
  to find dirty nodes more quickly~\cite{wbe,tali-garseil}.
This algorithm adds summary bits
  to identify and skip subtrees without any dirty elements.
While this reduces the search time,
  Double Dirty Bit still has to traverse the tree,
  starting from the root, to find dirty elements,
  and thus accesses not only the actually-dirty elements
  but many extra ``auxiliary nodes''.
On large pages,
  auxiliary nodes can significantly outnumber
  the actually-dirty elements;
  and since each node access can cause a cache miss,
  simply traversing these auxiliary nodes,
  just to check their summary bits,
  can stall the layout algorithm for hundreds of microseconds.
This problem is widely observed in practice;
  Google's widely-used web performance tool, Lighthouse,
  measures tree depth and maximum children count
  precisely because these parameters
  determine the number of auxiliary nodes.

\paragraph{Spineless Traversal}

We introduce \textit{Spineless Traversal}:
  a new, faster algorithm for incremental layout.
Unlike Double Dirty Bit,
  Spineless Traversal accesses only dirty elements,
  not auxiliary nodes.
It therefore reduces cache misses.
Spineless Traversal works by
  storing the set of dirty elements in a queue
  and jumping directly from one to the next,
  with no auxiliary nodes in between.

The key challenge is traversing the dirty nodes in the correct order. 
Recomputing a field on one node
  can mark fields on other nodes dirty,
  and the set of transitive dependencies is complex.
Fields must therefore be recomputed in a specific order,
  and Spineless Traversal must respect that order
  as it jumps from node to node.
Spineless Traversal thus stores
  a timestamp on each node,
  and uses a priority queue to traverse nodes
  in timestamp order.
To maintain timestamps as nodes are added and removed,
  Spineless Traversal \emph{order maintenance},
  which compute relative timestamps in a flexible way.
Both the priority queue and order maintenance structure
  are heavily optimized to make them competitive with Double Dirty Bit.

\paragraph{Implementation}

We implement Spineless Traversal in Megatron,
  a new compiler for incremental layout algorithms.
Megatron implements decades of research on attribute grammars:
  it statically analyzes the dependency graph,
  synthesizes recomputation and marking functions,
  and guarantees a correct incremental layout.
It implements standard optimizations
  (unboxing, interning, field packing, and jump tables)
  and compiles layout algorithms to highly efficient C++ code.
Megatron supports both Double Dirty Bit and Spineless Traversal
  via a common invalidation traversal interface,
  allowing an apples-to-apples comparison between them.

We compare the two algorithms on
  a significant fragment of web layout that includes
  line breaking, flex-box, intrinsic sizes, and many other features,
  benchmarking \TotalTraceCount real-world web pages
  like Twitter, Discord, Github, and Lichess.
Across \TotalDiffCount frames,
  Spineless Traversal is \DBPQtotal times faster on average.
Speedups are concentrated in the most latency-critical frames:
  on the \PctSmall of frames
  where at most 1\% of fields are recomputed,
  Spineless Traversal achieves a speedup
  of \MeanSpeedupSmall.

\section{Web Layout Background}

Complex
  web (Facebook, Amazon, GMail),
  desktop (VS Code, Figma, Slack),
  and mobile (WeChat) applications
  involve two software components:
  the application itself and the browser.
The application-level code, typically in JavaScript,
  implements the actual application logic and
  interacts with the browser by making DOM API calls
  that modify the browser's internal representation of the web page.
The browser then executes its ``rendering pipeline''---%
  event handling, hit testing, matching, styling,
  layout, paint, layerizing, rastering, and drawing---%
  which reads this internal representation and transforms it,
  step by step, into pixels on the screen.
Most execution time is typically spent in application code,
  but that code's effects are ultimately visible to the user
  only through ``system calls'' serviced by browser-level code.
For the application to be responsive and smooth,
  those calls have to be serviced
  in 16 milliseconds (60 frames per second).

Concretely, a browser application is written in HTML,
  which defines a tree called the ``document'',
  whose nodes include text, buttons,
  and structural elements like \texttt{div}s.
The application code binds specific callback functions
  to user actions like clicking on a button,%
  \footnote{Callbacks can also run in response to
    timers, network requests, or a dizzying array of
    other events.}
  and these callbacks can modify the HTML using DOM methods
  like \texttt{appendChild} or \texttt{setAttribute}.
When the callback finishes executing,%
\footnote{Or after multiple callbacks finish executing,
  as decided by the browser's task scheduler.}
  the browser executes the full rendering pipeline
  to display the results to the user.
Also, some DOM methods like
  \texttt{getBoundingRect} or \texttt{offsetX}
  read state computed during the rendering pipeline;
  calling these methods can requires the browser
  to perform additional passes through the rendering pipeline.
This is necessary for some common interactions like tooltips.

\subsection{The Layout Phase}
\begin{figure}
\includegraphics[scale=0.3]{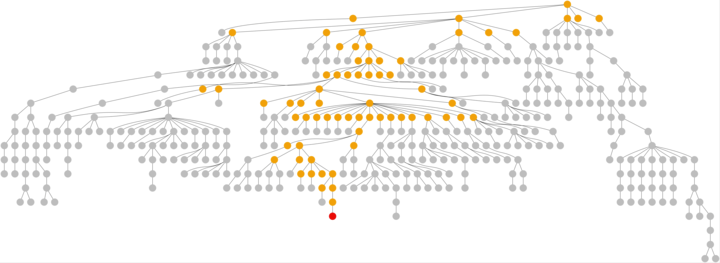}
\caption{The DOM tree for \texttt{google.com}, with 842 total nodes, a maximum fanout of 16, and a maximum depth of 18. The node marked red is part of the auto-complete suggestion box. The node color corresponds to the Double Dirty Bit algorithm, with gold representing auxiliary nodes and gray representing skipped nodes. Auxiliary nodes are a large fraction of the page even when only one node is (transitively) dirtied.
}
\label{fig:google}
\end{figure}

This paper is specifically concerned with the layout phase,
  a long-term focus of the programming languages community.
The layout phase traverses an intermediate structure
  called the ``layout tree'' and computes, for each node,
  a set of ``layout fields''
  including each element's size and position.
This computation proceeds in several passes,
  first computing intermediate layout fields
  like intrinsic width and height
  and current line ascent/descent
  before computing size and position.

The layout tree is basically the HTML tree,
  with minor differences for
  ``generated content'' (like bullets for list items)
  and ``fragmentation'' (like line breaking)
  that are not critical to this paper.
The trees are both big and unbalanced;
  the famously minimal Google home page page, for example,
  has 842 nodes, with a maximum fanout of 16 and depth of 18;
  it is drawn in \Cref{fig:google}.
In memory, the layout tree is stored as a pointer tree,
  with the children of each node stored in a doubly-linked list.%
\footnote{
This ensures that node insertions and deletions are fast,
  even in poorly-balanced trees.
}
The application can add or remove nodes from this tree,
  or write to their ``properties'' and ``attributes'',
  to update what the user sees.

Each node's layout fields depend on the layout fields
  of its neighbors.
For example, imagine a paragraph containing several lines;
  there would be a layout node for the paragraph
  and another for each line as children of the paragraph's.
The height of the paragraph, in this case,
  would be the sum of the heights of all its children
  plus any gaps between them,
  while a line's width would be its parent's width,
  minus some padding.
Of course, real-world web layout is much more complex;
  Chrome's implementation is about 100,000 lines of code
  and computes hundreds of fields per node.
To compute all these fields correctly, the layout algorithm
  recursively traverses the tree multiple times.
For example, the layout algorithm might first compute
  the intrinsic size of each element in a bottom-up traversal;
  then compute preferred sizes, in a top-down traversal;
  and then apply flexible sizing rules in a bottom-up traversal
  to finally compute each element's actual size.
Note that not only are multiple passes necessary,
  but that each pass must visit the nodes of the layout tree
  in a specific order so that each layout field's dependencies
  are satisfied.

\begin{figure}
    \centering
    \includegraphics[width=\linewidth]{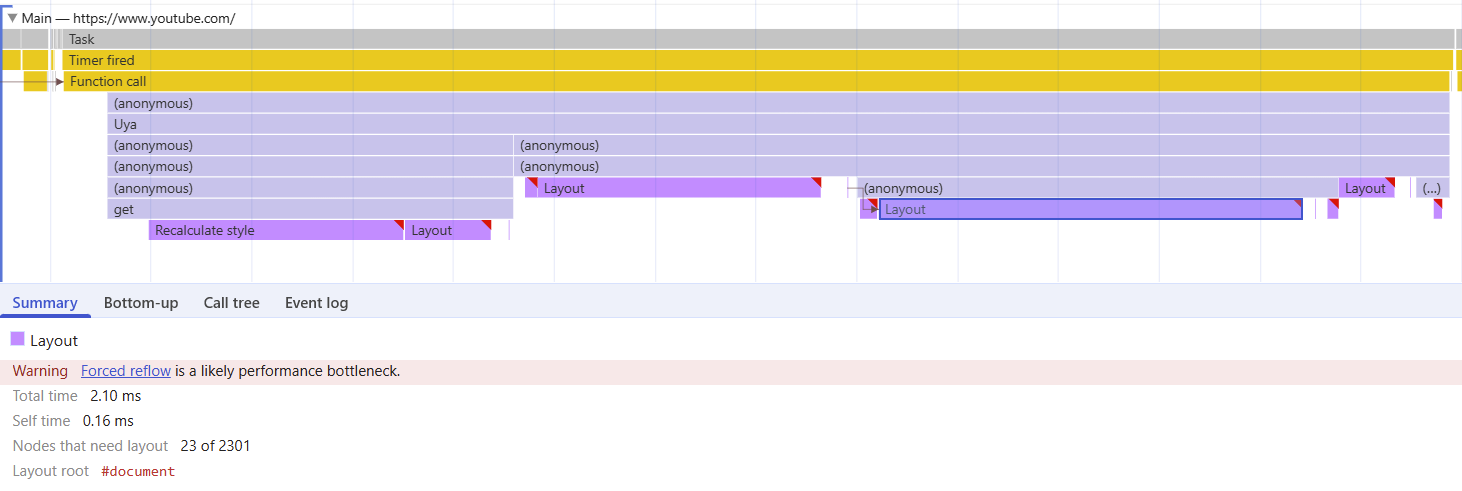}
    \caption{
      A trace of nine milliseconds of Chrome opening Youtube;
        time flows left to right while the call stack grows down.
      The violet ``recalculate style'' and ``layout'' blocks
        are phases of the browser rendering pipeline,
        with layout in total consuming four milliseconds;
        the faint vertical lines are half-millisecond marks.
      Layout happens multiple times in one frame
        due to ``forced reflows''.
      The ``Summary'' tab at the bottom shows
        that the selected (longest) layout
        only updates 23 of 2\thinspace301~nodes
        in the layout tree.}
    \label{fig:profile}
\end{figure}

\Cref{fig:profile} shows layout performance in practice:
  a trace of Google Chrome opening YouTube,
  captured and visualized using Chrome's ``Performance'' tools.
The trace covers 9~milliseconds of execution time,
  with time running left to right.
Function calls grow downward, and the leaves,
  labeled ``recalculate style'', ``layout'',
  and ``pre-paint'' (barely visible)
  are different phases of the rendering pipeline;
  the application-level code is offscreen to the left.
In these nine milliseconds,
  the application requests layout four times
  by calling APIs that require (``forced reflow'')
  multiple passes through the rendering pipeline.
These four layouts each take longer than
  the browser developers' target latency.

Below the trace, the ``Summary'' tab shows
  more information about a single, selected layout
  taking 2.1~milliseconds.
This layout operates
  on a layout tree of 2\thinspace301~nodes,
  starting from the root of the web page (\texttt{\#document}).
However, it only needed to update 23~of~them;
  in other words, this layout was incremental.
The long running time of the layout phase
  nonetheless caused unacceptable latency;
  in fact, this frame ``drops'',
  meaning the browser isn't able to update the page
  in time to show smooth animations and interactions.

\subsection{Formal Modeling}

Luckily, the programming languages community
  has developed a sophisticated understanding of layout.
Early work by Meyerovich and others%
  ~\cite{meyerovich-1,meyerovich-2,meyerovich-3}
  developed ``attribute grammars'' as a formalism
  for defining layout rules and implementations.
The later Cassius project~\cite{cassius-1,cassius-2,cassius-3}
  developed a full, standards-compliant implementation
  of a significant fragment of web layout in this formalism,
  and the \textsc{Hecate} and \textsc{Medea}
  tools~\cite{yufeng-1,yufeng-2}
  have shown that automatic synthesis can be scaled
  to such fragments.  

\begin{figure}
\begin{align*}
\text{Layout} &\coloneq  \text{Rule}^+; \textbf{schedule}\:\text{Pass}_n^+ \\
\text{Rule} &\coloneq
  \mathbf{def}\:\text{Pass}_n()\:\{\:
    A^+;\:
    \mathsf{children}.\mathsf{forEach}(\text{Pass}_n);\:
    A^+;\:
  \} \\
A \in \text{Assignment} &\coloneq
  \text{self}.V \leftarrow T \\[4pt]
T \in \text{Term} &\coloneq
  \text{if}\ T\ \text{then}\ T\ \text{else}\ T \mid
  F(T^+) \mid
  N? \mid
  N.V \mid
  \mathsf{attribute}[V] \mid
  \mathsf{property}[V] \\
N \in \text{Neighbor} &\coloneq
  \mathsf{self} \mid \mathsf{prev} \mid
  \mathsf{next} \mid \mathsf{parent} \mid
  \mathsf{first} \mid \mathsf{last} \\[4pt]
V \in \text{Variable} &\coloneq \text{layout fields} \quad\quad
F \in \text{Function} \coloneq \text{primitive functions}
\end{align*}
\caption{
  A minimal DSL for defining web layout
    as a set (\textsf{rules}) of passes
    performed in a specific order (\textsf{schedule}).
  The syntax $P^+$ represents a sequence of non-terminal $P$.
  Passes are in-order traversals of the layout tree
    performing a sequence of assignments to local fields
    while accessing fields of the current node or its neighbors.
}
\label{fig:dsl}
\end{figure}

\Cref{fig:dsl} defines an attribute grammar.
An attribute grammar is defined by
  a set of passes (the rules)
  performed in a certain order (the schedule).
Each pass performs a recursive, in-order traversal of the tree,
  computing some fields pre-order and some fields post-order;
  every field is written to exactly once in exactly one pass.
For each field assignment $\mathsf{self}.V \gets T$,
  the expression $T$ can refer
  to fields of $\mathsf{self}$ or
  to fields of its $\mathsf{parent}$,
  $\mathsf{prev}$ and $\mathsf{next}$ sibling,
  or $\mathsf{first}$ and $\mathsf{last}$ child;
  expressions can also test whether a given neighbor
  exists ($\mathsf{N?}$).
Computations can also refer to
  attributes or properties of the current node
  using $\mathsf{attribute}[x]$ or $\mathsf{property}[x]$.%
\footnote{
HTML attributes and CSS properties
  use two different namespaces
  because some names, like \texttt{height},
  appear in both sets; there is no other semantic difference.
Other accessible properties,
  such as the tag name or image width and height
  are modeled in our implementation as special properties.
}
All computations besides field assignments are pure,
  there are no other loops or data structures,
  and the only field access allowed is to a node's neighbors.
Despite this, even complex layout features
  like flexible box layout are expressible in such a DSL.

\begin{figure}
\begin{minipage}[b]{0.68\linewidth}
\begin{align*}
& \mathbf{def}\:\text{Pass}_1()\:\{ \\
& \quad \mathsf{self}.W \gets
        \mathbf{if}\:\mathsf{parent}?\:
        \mathbf{then}\:\operatorname{max}(0, \mathsf{parent}.W - 10)\:
        \mathbf{else}\:50; \\
& \quad \mathsf{children}.\mathsf{forEach}(\text{Pass}_1); \\
& \quad \mathsf{self}.H \gets
        \mathbf{if}\:\mathsf{last}?\:
        \mathbf{then}\:\mathsf{last}.HA + 10\:
        \mathbf{else}\:\mathsf{self}.\text{attribute}[\mathsf{height}]; \\
& \quad \mathsf{self}.HA \gets
        \mathbf{if}\:\mathsf{prev}?\:
        \mathbf{then}\:\mathsf{prev}.HA + \mathsf{self}.H + 5\:
        \mathbf{else}\:\mathsf{self}.H; \\
& \} \\
& \mathbf{schedule}\:\text{Pass}_1
\end{align*}
\caption{
  A minimal paragraph layout implementation,
    computing width $W$ and height $H$,
    with 5 pixels padding and 5 pixel gaps between lines.
  The intermediate $HA$ field sums the height
    of a node and all its previous siblings and gaps.
  This simple layout algorithm has one pass,
    but real-world layouts contain multiple.
}
\label{fig:layout-simple}
\end{minipage}\hfill%
\begin{minipage}[b]{0.28\linewidth}
\centering
\includegraphics[width=\linewidth,trim=0 0 11in 0,clip]{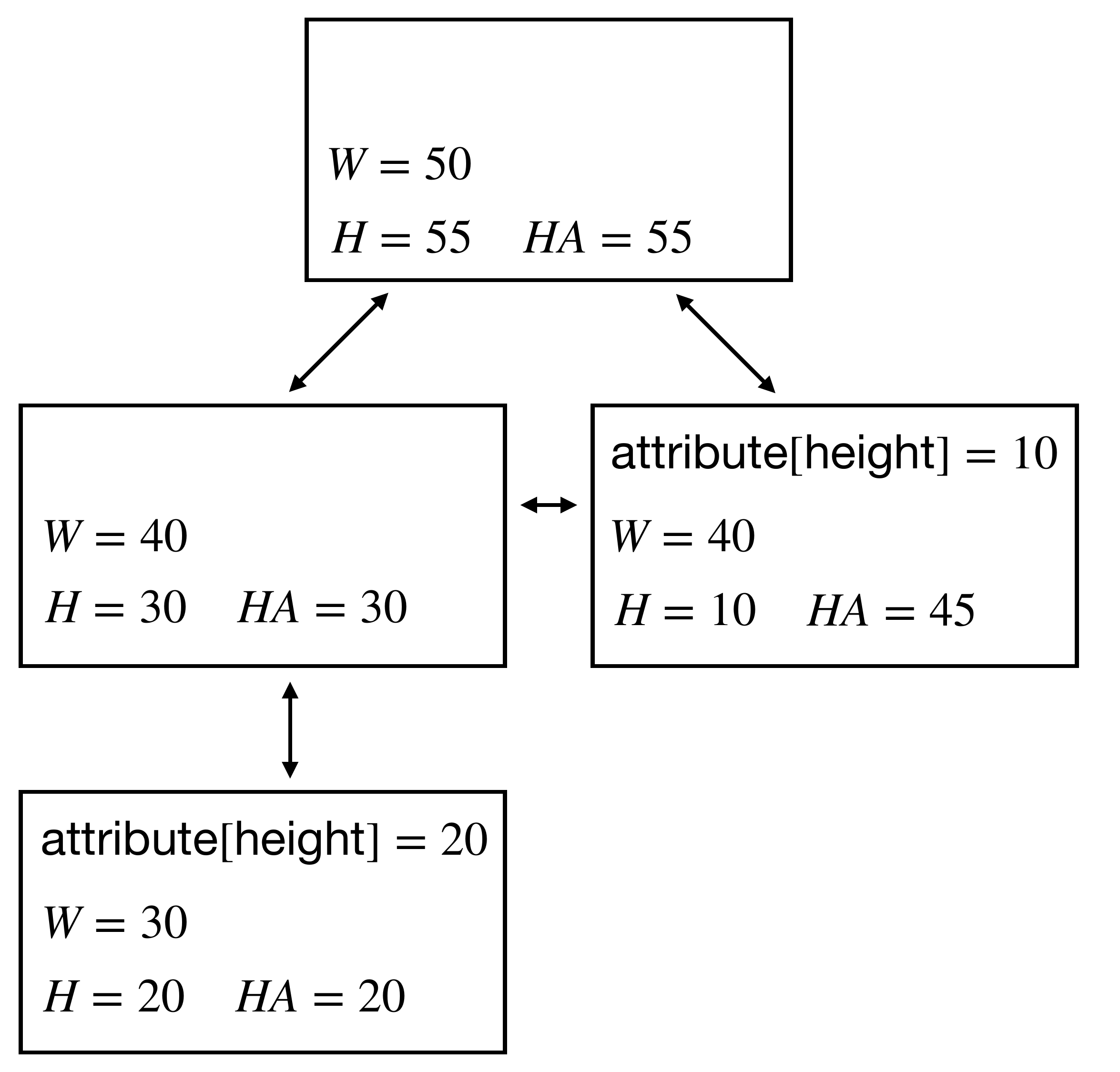}
\caption{The layout algorithm running on a layout tree of size 4. All nodes have an height attribute of 10.
}
\end{minipage}
\end{figure}
\Cref{fig:layout-simple} shows
  example layout rules for paragraphs and lines,
  defined using an attribute grammar.
Here the nodes for paragraphs and lines have three layout fields:
  a width $W$, a height $H$, and an intermediate field $HA$
  that computes the height of a node and its previous siblings
  (plus any gaps between lines).
Width information propagates from parents to children,
  starting at 50 pixels and subtracting, at each level,
  5 pixels of padding on the left and right.
Height information propagates from children to parents---%
  a node's height is the sum of its children's heights,
  plus 5-pixel gaps between lines---%
  and relies on the intermediate $HA$ field
  which propagates information from previous to next siblings.
Note that the idea of ``summing all children's heights''
  is not expressed with a loop;
  instead, it is expressed as the additional $HA$ layout field,
  which has its own computation rule.

Attribute grammar DSLs can be heavily optimized.
The layout fields can stored in the node itself,
  tightly packed to ensure locality.
Computations can use primitive data types
  like integers, floats, and enumerations;
  strings can be interned and hash tables statically flattened.
The only pointer accesses are to neighboring nodes,
  which are likely to be in cache.
Branch mis-prediction are relatively rare
  given the minimal control flow.
The tree structure itself does not change \emph{during layout}.
This efficiency is critical to browser developers.

A key property of attribute grammar DSLs like this one
  is that dependencies and traversal orders are static.
Specifically, for any field assignment,
  examining the expression $T$ reveals
  which other fields on which other nodes it depends on.
These static dependencies are critical for invalidation.
Also, the schedule language ensures that
  the relative order in which two fields are computed
  never changes, even as nodes are added or removed.
This is critical for Spineless Traversal.
In this paper, we assume
  that the rules have no cyclic dependencies
  and that the schedule
  respects field dependencies;
  we also assume that the schedule has already been optimized
  by fusing traversals and ordering field assignments.
A substantial literature exists on these topics~%
  \cite{grafter,yufeng-1,yufeng-2};
  the layout implementation in our evaluation
  is detailed in Section~\ref{sec:layout-impl}.
In any case, while this paper focuses on web layout,
  we expect Spineless Traversal to be applicable
  to incremental computations in other domains,
  including in compilation, static analysis,
  computer graphics, and databases,
  as long as they can be expressed as an attribute grammar
  in this or a similar DSL.
\section{Incremental Layout}

An \emph{incremental} layout algorithm
  reuses layout field values between layouts when possible.
For example, when the user moves their mouse,
  a tooltip may need to move to follow the cursor;
  incremental layout would re-compute the position of the tooltip
  while reusing all other layout fields for all other elements.
Most layouts change only a small fraction of fields---%
  especially the most latency-critical interactions
  like hovers, drags, animations, and text editing---%
  so incremental layout can be dramatically faster
  than computing each layout from scratch.

\subsection{Dirty Bits}
\label{sec:recompute-phase}

Incremental layout is conceptually straightforward. 
Each layout field has a corresponding \textit{dirty} bit,
  which defines whether that layout field needs to be recomputed.%
\footnote{\emph{Packed} layout fields can all have
  the same corresponding dirty bit,
  which is set if any of those fields need to be recomputed.}
APIs that write to an attribute or property,
  or add or remove layout nodes,
  set the dirty bits for all layout fields
  that read from those fields or node pointers.
For example,
  in the paragraph layout of \Cref{fig:layout-simple},
  when a node's \textsf{height} attribute is changed,
  its $H$ field must be marked dirty.
When a subtree is deleted,
  its next sibling's $HA$ field must be marked dirty.
If the subtree was the last child of its parent,
  the parent's $H$ field must also be marked dirty.
This set of fields can be statically determined in our DSL
  and in fact our Megatron compiler synthesizes
  this marking code automatically.

Incremental layout then has the task of clearing dirty bits
  by recomputing dirty fields.
Conceptually this is simple:
  find all dirty bits on all nodes and recompute
  the fields they correspond to.
However, when a layout field is recomputed,
  its value might change,
  and then any layout fields that read the old value
  are would be out of date.
Thus, when incremental layout modifies a layout field,
  it must dirty all layout fields that depend on it,
  ``propagating'' dirty bits through the document.
For example,
  in the paragraph layout of \Cref{fig:layout-simple},
  when a node's $H$ field is recomputed,
  its $HA$ field must be marked dirty.
When that $HA$ field is recomputed,
  its next sibling's $HA$ field is then marked dirty in turn.
Because of dirty bit propagation,
  the set of layout fields \emph{currently} marked dirty
  is distinct from the set of all layout fields
  that incremental layout must ultimately recompute.

This second set is \emph{discovered}
  in the process of performing incremental layout:
  dirty bits are only propagated
  when a layout field's value changes,
  which is only known when that field is recomputed
  and the old and new value can be compared.
For example, suppose the user is typing out a paragraph of text,
  laid out using \Cref{fig:layout-simple}'s layout algorithm.
Most edits just add text to a single line,
  not affecting its height or available width;
  these only end up dirtying a single field on a single node.
This is why incremental layout works:
  many interactions---%
  especially latency-critical ones like
  hovers, drags, animations, and text editing---%
  do not change fields on containing elements,
  and end up affecting only a small portion of the page.

\subsection{Recomputation Order}

Incremental layout strives to minimize
  the number of layout fields recomputed.
That requires ensuring that, once recomputed,
  a layout field is not marked dirty again during that layout.
Luckily, there is a simple way to guarantee this:
  a correct from-scratch layout
  always processes a field's dependencies before the field itself,
  so recomputing layout fields
  in the \emph{same relative order} as a from-scratch layout
  ensures that each layout field is marked dirty at most once.
A naive incremental layout algorithm might thus
  execute the same layout schedule as a from-scratch layout
  but skip re-computing any fields that aren't dirty.
This algorithm recomputes the minimal number of layout fields,
  clears all the dirty bits,
  and avoids ever recomputing a layout field more than once.

However, this naive algorithm isn't particularly fast:
  it must access every layout node to check its dirty bits,
  which incurs a lot of cache misses.
In the common case, most dirty bits aren't set;
  these ``auxiliary'' accesses cause needless cache misses
  and end up wasting a significant fraction of run time.
For incremental layout to be fast,
  auxiliary accesses---%
  accesses that don't result in field recomputations---%
  must be minimized.

\subsection{The Double Dirty Bit Algorithm}

The state-of-the-art Double Dirty Bit algorithm
  uses summary bits to reduce auxiliary accesses;
  it is used, with variations, in all major rendering engines.
For each dirty bit, we add a second ``summary bit'',
  which is set for any node where the corresponding dirty bit
  is set anywhere in the \emph{subtree} rooted at that node.
When a field is marked dirty, its dirty bit is set,
  and the associated summary bit is also set
  on that node and all its ancestors.
When performing incremental layout,
  subtrees whose summary bits are clear can be skipped.
Figure~\ref{fig:find-dirty-nodes} contains
  pseudo-code for Double Dirty Bit
  as an iterator over dirty nodes.

\begin{figure}
\begin{minipage}[b]{0.5\linewidth}
\begin{verbatim}
def mark_dirty(self):
    self.dirty_bit = True
    self.set_summary_bit()

def set_summary_bit(self):
    if self.summary_bit: return
    self.summary_bit = True
    if self.parent:
        self.parent.set_summary_bit()
\end{verbatim}
\caption{Setting the summary bit for a node.}
\label{fig:set-summary-bits}
\end{minipage}\hfill%
\begin{minipage}[b]{0.5\linewidth}
\begin{verbatim}
def find_dirty_nodes(self):
    if self.dirty_bit:
        yield self
        self.dirty_bit = False
    if self.summary_bit:
        # Access auxiliary nodes
        for child in self.children:
            find_dirty_nodes(child)
        self.summary_bit = False
\end{verbatim}
\caption{Finding the dirty nodes in a tree.}
\label{fig:find-dirty-nodes}
\end{minipage}
\end{figure}

Skipping subtrees with no dirty bits
  reduces the number of auxiliary accesses
  and greatly improves performance~\cite{tali-garseil,wbe}.
However, \emph{some} auxiliary accesses remain.
Specifically, when a dirty bit is set on some node,
  summary bits will be set for all ancestors of that node,
  which Double Dirty Bit will have to check.
Moreover, for each ancestor,
  Double Dirty Bit will recurse into its children,
  adding even more auxiliary accesses.
Since layout trees are poorly balanced and often very wide,
  the number of auxiliary accesses can be large.
For example,
  a single dirtied node in \Cref{fig:google} has 66 auxiliary nodes,
  and on larger pages the dirty-to-auxiliary ratio can be even worse.

Developers can sometimes reduce the number of auxiliary accesses
  by reorganizing large, complex pages;
  performance monitoring tools
  like Google's Lighthouse~\cite{lighthouse}
  will suggest doing so.
But this can require global changes to the shape of the layout tree,
  which (in modern frameworks like React)
  requires refactoring the application as a whole.
Naturally, an invalidation algorithm that simply
  did not require so many auxiliary nodes
  would be a superior solution.
And while this is a quite specialized performance problem,
  it is a major source of latency in existing web browsers,
  which in turn are both critical application platforms
  and also already highly optimized,
  meaning remaining sources of latency are
  particularly challenging.
\section{Spineless Traversal}

Spineless Traversal improves on Double Dirty Bit
  by jumping directly between dirty nodes
  without accessing any auxiliary nodes;
  as a result, it suffers dramatically fewer cache misses.
Achieving this requires a more computationally heavy approach:
  storing all dirty nodes in a priority queue
  and maintaining the correct traversal order
  using an order maintenance data structure.
Spineless Traversal's savings in cache misses
  typically outweigh the greater computational requirements
  of these data structures.
Since Spineless Traversal is complex,
  this section develops it incrementally,
  first introducing the idea of a queue storing dirty nodes,
  then adding timestamps to maintain traversal order,
  and finally introducing the order maintenance structure
  to handle node insertion and deletion.
A final subsection optimizes bulk insertions and deletions.

\subsection{Jumping Directly to Dirty Nodes}

To jump directly to dirty nodes,
  we introduce a queue of pointers to dirty nodes.
More specifically, elements of the queue represent
  dirty bits in the layout tree that need to be cleared,
  and are represented in the queue by
  a pointer to the relevant layout node
  and an enumeration identifying
  which dirty bit needs to be cleared.
When a dirty bit is set,
  the corresponding node and enumeration are added to the queue;
  the order of nodes in the queue will be explained later.
Note that dirty bits are still present on layout nodes;
  elements are only added to the queue
  when a dirty bit is set for the first time,
  so the queue does not contain duplicates.
To perform an incremental layout,
  elements are popped from the queue
  and the relevant fields on the relevant node are recomputed,
  clearing the relevant dirty bit.
In the process, new dirty bits may be set,
  so new elements may be added to the queue.
This is repeated until the queue is empty.
Only dirty nodes are ever in the queue,
  meaning only they---not auxiliary nodes---are ever accessed.

Concretely, the queue is stored in a packed array,
  which due to heavy use will remain in cache.
Clearing a dirty bit thus suffers cache misses
  only for the dirty element,
  plus its at most five neighbors.
This means clearing a dirty bit incurs
  a maximum of six L2 cache misses---%
  much fewer than the number of auxiliary accesses
  typical of Double Dirty Bit.

\subsection{Maintaining Queue Order}

To ensure that each field is only recomputed once,
  queue elements must be in the right order.
We thus add a timestamp field
  for every dirty bit on every node,
  giving timestamps the same relative order
  as in a from-scratch layout.
In the full Spineless Traversal algorithm,
  as explained below,
  this timestamp is an ``order maintenance object'',
  but for now the reader can imagine it an integer counting up from 0.
The queue of dirty nodes is then refined to a priority queue,
  ordered by these timestamps.
The priority queue ``pops'' the lowest timestamp first,
  so Spineless Traversal clears dirty bits in timestamp order
  and thus clears each dirty bit exactly once.

Concretely, we use a min-heap as our priority queue,
  which is cache-friendly and requires
  relatively few operations for each push and pop.
Timestamps are stored adjacent to dirty bits,
  meaning they do not introduce any new L2 cache misses.
The priority queue is typically small:
  while a web page may have thousands of nodes,
  with each node having dozens of fields in our evaluation,
  the priority queue typically contains less than 1000 elements,
  and for the most latency-critical interactions,
  like hovers or drags, it can contain 100 or fewer.
With such a small size, a priority queue push/pop requires
  5--10 timestamp comparisons,
  which can be performed in roughly the time
  of one to three L2 cache misses
  in our optimized implementation.

\subsection{Order Maintenance}

The final challenge is efficiently assigning timestamps
  to every dirty bit on every node.
Simple incrementing integer timestamps, sadly, don't work:
  inserting a node would requiring adjusting all later timestamps,
  which would introduce its own set of auxiliary accesses.
Instead, following SAC~\cite{SAC},
  Spineless Traversal uses
  an \emph{order maintenance} data structure (OM)
  to assign timestamps.
First introduced by \citet{OM},
  order maintenance maintains a totally ordered set of objects
  while allowing objects
  to be added and removed from the order arbitrarily.
Crucially, adding, removing, and comparing nodes takes $O(1)$ time.
Abstractly, order maintenance provides the following API:

\begin{enumerate}
\setlength{\itemindent}{8em}
  
\item[$\mathsf{Compare}(p, q)$] Decides whether $p$ or $q$ comes first in the order (or are equal).
\item[$\mathsf{Head}()$] Returns the first object of the order.
\item[$\mathsf{Create}(p)$] Creates and returns a new object right after $p$.
\end{enumerate}

\noindent
Deleting OM objects is also possible, though by default
  our implementation does not do so.%
\footnote{On long-running pages
  this causes very-slowly-growing memory usage.
  Enabling OM object deletion avoids this but adds
  roughly 2\% to Spineless Traversal's running time.}

Our implementation is based on that by \citet{SOM},
  which uses a two-level structure with
  a doubly-linked list of doubly-linked lists
  (\Cref{fig:om}).
Objects are represented by nodes in the lower-level lists.
Both levels are ordered;
  the total order is lexicographic,
  traversing higher-level lists in order
  and then, for each higher-level list,
  its lower-level list in order.
Each object (node in the lower-level list)
  maintains a pointer to its higher-level list cell;
  two objects are in the same low-level list
  if they have the same higher-level pointer.
To allow fast comparisons between nodes,
  each low-level and high-level list cells stores
  an unsigned integer of fixed size
  (in our implementation, 32~bits)
  called its label.
Within a list, node labels are strictly increasing;
  this makes comparisons fast.
Specifically, comparison has two cases:
  if the two objects are in the same low-level list,
  their labels are compared directly,
  while if they are in in different ones,
  their parents' labels are compared.
This comparison operation is
  the bulk of the Spineless Traversal time
  so its speed is essential;
  \Cref{sec:opt} discusses critical micro-optimizations
  that bring its latency down to around 5~cycles.

To create an object inside an order maintenance structure,
  a new lower-level list cell is created
  whose label is the average of the two neighboring labels.%
\footnote{
  When creating a node after the last node,
  the maximum representable number is used as the larger number.
}
If the two labels differ by exactly 1, however,
  this would repeat a label.
In this case, the data structure re-balances itself,
  evenly reassigning labels to existing objects.
This process might
  create a new higher-level list cell
  to split a lower-level list in two,
  ensuring a sufficiently large gap between its cells.
Rebalancing is algorithmically tricky
  but is not a significant time sink in our use case,
  so we do not detail re-balancing here;
  interested readers can find a description in \citet{SOM}.

\begin{figure}
\begin{tikzpicture}
\draw[blue, very thick] (1.5,0) rectangle (3.5,1) node[pos=.5]{10};
\draw[ultra thick, <->] (3.5,0.5) -- (7.5,0.5);
\draw[blue, very thick] (7.5,0) rectangle (9.5,1) node[pos=.5]{20};

\draw[orange, very thick] (0, -2) rectangle (1,-1) node[pos=.5]{10};
\draw[very thick, <->] (0.5,-1) -- (2,0);
\draw[very thick, <->] (1,-1.5) -- (2,-1.5);
\draw[orange, very thick] (2, -2) rectangle (3,-1) node[pos=.5]{15};
\draw[very thick, ->] (2.5,-1) -- (2.5,0);
\draw[very thick, <->] (3,-1.5) -- (4,-1.5);
\draw[orange, very thick] (4, -2) rectangle (5,-1) node[pos=.5]{20};
\draw[very thick, ->] (4.5,-1) -- (3,0);

\draw[orange, very thick] (6, -2) rectangle (7,-1) node[pos=.5]{10};
\draw[very thick, <->] (6.5,-1) -- (8,0);
\draw[very thick, <->] (7,-1.5) -- (8,-1.5);
\draw[orange, very thick] (8,-2) rectangle (9,-1) node[pos=.5]{15};
\draw[very thick, ->] (8.5,-1) -- (8.5,0);
\draw[very thick, <->] (9,-1.5) -- (10,-1.5);
\draw[orange, very thick] (10, -2) rectangle (11,-1) node[pos=.5]{20};
\draw[very thick, ->] (10.5,-1) -- (9,0);

\end{tikzpicture}
\caption{An Order Maintenance data structure. The blue node represent the higher level doubly linked list, and each node store a lower level doubly linked list, denoted by the orange node. Lower level node also store a pointer to the higher level node. Each node additionally hold an unsigned integer, label, such that inside a single list the node earlier have a strictly smaller label then the node later.}
\label{fig:om}
\end{figure}

Concretely, the timestamp for each dirty bit
  is now represented by a pointer to the lower-level OM object.
Priority queue elements are a node pointer,
  an enumeration naming the dirty bit,
  and a pointer to the OM node,
  padded to a total of 16 bytes to align with cache lines.
To compare two timestamps, both pointers must be dereferenced
  to access the two lower-level OM objects.
The associated higher-level OM objects must also be accessed.
In the typical case, where the priority queue remains small,
  all of the OM objects in the priority queue
  are already in cache,
  so ``pops'' have no additional cache misses,
  while ``pushes'' add just one or two, for the new element OM cells.
In total, all the priority queue and order maintenance operations
  thus add the equivalent of 3--6~L2~cache misses in latency,
  about 100--300 cycles.
Since recomputing a field, accessing neighbors' fields,
  and setting neighbors' dirty bits can itself take hundreds of cycles,
  Spineless Traversal's overhead is fairly small.

\subsection{Subtree Insertion}
\label{sec:tree-insertion}

Bulk insertions into the layout tree
  are common in ``lazy loading'' patterns:
  a ``shell'' web page loads first and shows a loading indicator;
  then the ``content'' loads and is inserted as a large subtree,
  replacing the loading indicator.
This pattern is encouraged by frameworks like React,
  and can occur in several stages, with a ``shell''
  first inserting ``subshells'' which
  themselves load subcomponents in turn.
Efficiently handling these bulk insertions
  requires special care in Spineless Traversal,
  as bulk insertions are typically responsible
  for Spineless Traversal's worst performance
  relative to Double Dirty Bit.

The basic issue is that when a fresh subtree is inserted,
  every field needs to be computed
  and every OM object needs to be initialized,
  all without causing the priority queue to grow too large.
Our solution add is to add ``initialization passes''
  as special elements in the priority queue:
  besides $(v, n)$ pairs for a dirty field $v$ on node $n$,
  an element can also be $(p, r)$,
  a pass $p$ that needs to initialize the entire subtree rooted at $r$.
The timestamp for such a special element
  is just after the last field assigned by $p$
  in $T$'s previous sibling or parent.%
\footnote{
  An edge case is inserting a subtree into
    a subtree that itself has not yet been laid out.
  In this case no further actions need to be taken,
    since both subtrees will be visited together.}

When one of these $(p, r)$ elements is popped from the queue,
  the pass $p$ is performed on the whole subtree under $r$,
  creating all necessary order maintenance nodes
  and computing all fields assigned by $p$.
Since all data accesses are local,
  no existing nodes will refer to any newly-inserted node
  except the root node,
  so no nodes need to be dirtied when running $p$ on the subtree.
This means that, when initializing a subtree,
  no priority queue operations or dirty bit propagations
  need to be performed,
  which makes subtree insertion faster.
However, order maintenance objects still need to be created 
  for every node in the subtree,
  so Spineless Traversal is still
  typically slower than Double Dirty Bit for subtree insertion.

When deleting a subtree,
  some nodes in the subtree may
  already be in the priority queue,
  like when a subtree is inserted and then deleted.
To avoid unnecessary recomputation,
  we add a ``deleted'' bit to each node
  and skip recomputing fields on deleted node.

\section{Megatron}

To evaluate Spineless Traversal,
  we developed an attribute grammar engine called Megatron
  which implements the DSL of \Cref{fig:dsl}.

\subsection{Compiler Interface}
\label{sec:compiler}

The Megatron compiler
  parses layout rules and schedules
  and compiles them to a fast incremental C++ layout algorithm.
Concretely, the output of Megatron
  is a \texttt{recompute} function,
  which is passed a node and a field name
  and which recomputes that field on that node,
  setting dirty bits for any dependent fields.
To produce this output, the compiler
  analyzes the static dependencies of each field assignment,
  synthesizes dirty bit propagation code,
  and generates packed, cache-friendly data structures.

To evaluate performance,
  we link the generated \texttt{recompute} function
  to a driver program that parses and replays web layout traces
  captured from a real web browser.
The trace is structured as a set of frames,
  which each contain some number of modification actions,
  after which incremental layout is performed.
The execution time of all of these steps is measured
  at very high precision using \texttt{rdtsc}.
Specifically, the driver program
  separates ``overhead'' time---%
  including setting dirty or summary bits,
  allocating OM nodes, and pushing to the priority queue---%
  from ``evaluation'' time,
  which includes the field recomputation itself.%
\footnote{
  Other parts of the driver program,
    like reading and parsing traces, are not measured
    because they are the same for all layout algorithms.}
Evaluation time is nearly identical
  across different invalidation algorithms,
  but overhead time differs dramatically, as expected.

\begin{figure}[tbp]
\begin{verbatim}
val dirty : node -> field_name -> unit
val clean : root : node -> recompute : (node -> field_name -> unit) -> unit
\end{verbatim}
\caption{
  The ``invalidation traversal'' API.
  The \texttt{dirty} method sets the dirty bit
    corresponding to a given field on a given node,
    while the \texttt{clean} function
    invokes the \texttt{recompute} function
    on every dirty field in the layout tree.
  In the actual implementation,
    the node and unit types are staged,
    but the \texttt{field_name} isn't,
    to maximize the performance of the generated code.}
\label{fig:traversal-api}
\end{figure}

To allow for a head-to-head comparison
  between Double Dirty Bit and Spineless Traversal,
  the driver program is parameterized over
  a simple ``invalidation traversal'' API.
It has two key methods:
  a \texttt{dirty} method
  that marks a node's field for later re-computation
  and a \texttt{clean} method
  that invokes \texttt{recompute} on all dirty fields.
Signatures for both functions are shown
  in \Cref{fig:traversal-api};
  there are also boilerplate methods
  to initialize data structures.
The \texttt{dirty} method is called
  both when modifying the layout tree
  and also from the \texttt{recompute} function itself,
  to dirty dependents when a field's value changes.
Three invalidation traversals are available in Megatron:
  a naive traversal that visits each node,
  Double Dirty Bit, and Spineless Traversal.

\subsection{Dependent Synthesis}

The most critical responsibility of the compiler
  is synthesizing code to sets dependent dirty bits
  every time a layout field is modified.
Megatron follows the standard approach from prior work.
In short, for every field $U$,
  the compiler identifies all fields on all nodes
  that are affected when the value of $U$ changes,
  and generates code to set their dirty bits
  by calling the \texttt{dirty} method.
To identify affected fields,
  the compiler examines every \emph{other} field assignment
  $\mathsf{self}.V \gets T$ in the layout rules.
For every field access $N.U$ in $T$,
  we know that $\mathsf{self}.V$ depends on $N.U$,
  meaning that any changes to $\mathsf{self}.U$
  must mark $N^{-1}.V$ as dirty;
  here $N^{-1}$ inverts the relation $N$ by
  flipping \textsf{next} and \textsf{previous},
  mapping \textsf{first} and \textsf{last} to \textsf{parent},
  and mapping \textsf{parent} to all children.

Insertions and deletions also set dirty bits.
Inserting a node changes the meaning
  of the \textsf{prev} and \textsf{next} pointers
  for its siblings, and possibly
  the \textsf{first} and \textsf{last} pointers
  for its parent;
  deleting a node does the same.
This means that inserting or deleting a node
  must dirty every field computation
  that reads \emph{any} field from those pointers,
  and also any field that uses $\mathsf{N?}$ expressions.
In Megatron, the analysis we perform is flow-insensitive;
  a real-world implementation might instead use
  a flow-sensitive analysis to dirty fewer nodes,
  but the engineering trade-offs are more challenging
  and we felt that the focus of this paper---%
  the invalidation traversal---%
  would have similar impact in either case.

As an example, consider the layout rules in \Cref{fig:layout-simple}:
\begin{enumerate}
\item The $W$ field depends on \textsf{parent?} and $\mathsf{parent}.W$.
As the \textsf{parent} of a node cannot change,
  the only 'real' dependency is on $\mathsf{parent}.W$.
Thus, setting $\mathsf{self}.W$ must dirty $\mathsf{parent}^{-1}.W$,
  meaning that after changing $\mathsf{self}.W$
  we must dirty all children's $W$ fields.
\item The $H$ field depends on $\mathsf{last}?$, $\mathsf{last}.HA$,
  and $\mathsf{self}.\text{attribute}[\mathsf{height}]$,
  and the $\mathsf{last}?$ dependency is subsumed by
  the $\mathsf{last}.HA$ dependency.
Taking inverses, modifying $\mathsf{self}.\text{attribute}[\mathsf{height}]$
  must dirty $\mathsf{self}.H$ and,
  if the node is a last child,
  modifying $\mathsf{self}.HA$ or inserting/deleting a node
  must dirty $\mathsf{parent}.H$.
\item The $HA$ field depends on $\mathsf{prev}?$, $\mathsf{prev}.HA$, 
  and $\mathsf{self}.H$, with the $\mathsf{prev}?$ dependency subsumed.
Taking inverses, 
  modifying $\mathsf{self}.H$ must dirty $\mathsf{self}.HA$%
\footnote{
As described later, field packing would likely assign
  the $H$ and $HA$ fields the same dirty bit;
  in this case, modifying $\mathsf{self}.H$
  need not dirty $\mathsf{self}.HA$, since it is already dirty.
}
  and,
  if the node is not a last child,
  modifying $\mathsf{self}.HA$ or inserting/deleting nodes
  must dirty $\mathsf{next}.HA$.
\end{enumerate}

These modification rules, gathered from analyzing each field access,
  are grouped by which field is being modified
  and then injected into the relevant case
  of the \texttt{recompute} function.
For example, the code to recompute $HA$
  would check whether or not the node is a last child
  and dirty either $\mathsf{parent}.H$ (if so)
  or $\mathsf{next}.HA$ (if not).
Additionally,
  Megatron makes sure to only dirty any given field
  once per field recomputation.
For example, if a field $N.V$ is used twice in an expression,
  or if two different accesses $\mathsf{first}.V$ and $\mathsf{last}.V$
  have the same inverse, the field is only dirtied once.
This deduplication is especially challenging
  in the case of $N?$ expressions,
  since whether or not a field is dirtied can depend
  on whether the node is the first or last child of its parent.
This deduplication is complex but, ultimately,
  possible to perform statically,
  which is critical to ensure maximal performance.

To ensure our implementation is correct,
  the compiler can also generate a from-scratch layout function
  which does not use dirty bits at all
  and instead recomputes the entire layout from scratch.
This was extremely valuable during development
  and gives us confidence that our invalidation algorithm is correct.

\subsection{Code Generation}

\begin{figure}[bt]
\begin{verbatim}
val evaluate : term -> env : (variable -> value) -> value
val evaluate_staged : term -> env : (variable -> value code) -> value code
\end{verbatim}
\caption{
  Converting an interpreter into a compiler by staging.
  Terms and variable names are static so are not staged;
    the initial environment and final outputs, however,
    are staged by transformation into an IR.}
\label{fig:stage}
\end{figure}

Megatron's code generation has three key steps:
  generating the node data structure,
  generating the field value computations,
  and generating the top-level \texttt{recompute} function.
To simplify code generation and optimization,
  the compiler is organized as a staged interpreter~\cite{MetaOCaml}.
That is, we first implemented an interpreter,
  and then added staging annotations that generate a custom IR;
  \Cref{fig:stage} summarizes the approach.
The custom IR is then converted to C++
  using traditional compiler optimizations.
This approach was critical,
  as we found compiler correctness challenging:
  flaws in the dependency analysis could have unintuitive,
  long-distance consequences on program behavior
  that were difficult to distinguish from incorrect optimizations.
The staging step allowed for much easier debugging.

\paragraph{Data structures}
To generate efficient node data structures,
  Megatron type-checks all fields, attributes, and properties
  using Hindley-Milner type inference~\cite{HM}
  and uses appropriate unboxed C++ member variables
  to store the relevant values.
Importantly, this means a single node and all its fields
  are contiguous in memory
  (as it would be a real web browser)
  with a minimum of pointers
  (beyond the standard parent/first/last/next/previous pointers
  to other layout nodes),
  with field access compiled to a memory offset.
All string values (like the keyword values for \texttt{display})
  are interned and represented in C++ as
  a single \texttt{enum} type,
  meaning that no string allocation or comparison
  is performed at runtime.
Discriminated unions are used for fields with units
  (for example $\text{property}[\mathsf{length}]$
    can be an absolute length or a percentage).
Dirty bits, summary bits, and timestamps
  are placed adjacent to the fields they cover.

To reduce overhead as much as possible,
  Megatron \emph{packs} multiple co-computed fields,
  covering them with a single dirty bit.
Packed fields are laid out adjacent in memory
  because they are written to, and often read from,
  in rapid succession.
While more sophisticated techniques exist~\cite{yufeng-2},
  Megatron simply uses two dirty bits
  for each in the layout rules:
  one for the pre-order-computed fields
  and one for post-order-computed fields.
The intuition is that this minimizes
  the number of unique dirty bits
  set during dirty bit propagation
  and reduces the size of the priority queue
  and the number of order maintenance objects needed.
Field packing has complex trade-offs
  not implemented in Megatron
  but common in real browsers;
  however, Spineless Traversal would work
  with any field packing.

\paragraph{Field computations}
To generate efficient field computations,
  the compiler applies strength reduction and dead code elimination,
  implemented as a simple walk over the AST.
The bottom line is that the compiled code is
  long but readable, idiomatic, and fairly low-level C++ code
  with no allocation, hash, or string operations.
The C++ compiler is then invoked, which performs its own optimizations.
This is critical because it means that,
  like in a real web browser,
  computing fields is very fast
  and the cache misses from finding dirty fields
  are a measurable fraction of the runtime.
The \texttt{recompute} function uses its field name argument
  to determine which field computations to run.
For Spineless Traversal,
  the field name is an \texttt{enum}
  and is used as an index into a jump table
  to perform the relevant recomputation;
  the special $(p, r)$ elements are included
  in the same \texttt{enum}.
For Double Dirty Bit, the field name is known
  entirely statically, and our compiler
  simply outputs the relevant code.
\section{Optimizations}
\label{sec:opt}

Maximizing Spinless Traversal's advantage
  over Double Dirty Bit
  requires careful optimizations
  to reduce work needed, improve cache locality,
  reduce memory traffic,
  and alleviate branch mispredictions.
Unlike the optimization passes of the Megatron compiler,
  which apply to all invalidation algorithms,
  these optimizations are implemented
  in the Spineless Traversal invalidation algorithm
  and only affect it.

\subsection{Queue Compression}

Although the queue size is typically small, often only tens of elements long, it can occasionally become very large. This typically happens when the entire page must be re-laid-out, such as when the browser window is resized. In this case, every dirty bit ends up in the priority queue, and the queue can grow to thousands of elements long. This not only means that more queue pushes and pops are needed, but also that those pushes and pops take longer.

In theses cases the queue is often dense: for every dirty bit in the queue, its successor is also likely dirty and enqueued as well. Leveraging this insight, we can shrink the queue by representing sequences of enqueued dirty bits implicitly. That is, every time a queue element is popped, Megatron not only recomputes those fields but it also checks if the succeeding dirty bit is dirty and, if so, recompute it too. Megatron continues recomputing successive dirty bits until it reaches a cleared dirty bit. When a dirty bit is set, it doesn't need to be enqueued if its predecessor is dirty. This leads to a significant reduction in queue length, and thus speeds up cases where many fields have to be recomputed, such as during initial loads and bulk inserts.

In principle, Megatron could even avoid checking predecessor dirty bits when this optimization is statically known to apply. However, we did not implement this further optimization.

\subsection{Pointer Compression and OM Allocation}
Order maintenance objects have to be allocated every time
  new layout nodes are inserted;
  optimizing that allocation is essential.
We use a hand-written pool allocator;
  in fact, we use separate pools
  for high- and low-level list cells
  to enhance locality.
Since allocations are always the same size,
  and since layout is single-threaded,%
\footnote{As it is in all major browsers.}
  our custom allocator is significantly
  simpler than the system \texttt{malloc}.

The allocator, \texttt{OMPool},
  is shown in Figure~\ref{fig:allocator}.
It is parameterized by \emph{two} types:
  the type of allocated object \texttt{T}
  and the index type \texttt{P} for pointers to allocated objects.
Crucially,
  \texttt{malloc} and \texttt{free} return and consume
  the pointer type \texttt{P} instead of raw pointers.
(The \texttt{addressof} function converts
  the pointer type \texttt{P} to a raw pointer
  so that it can be dereferenced.)
Making \texttt{P} a small integer type,
  like \texttt{uint32} or \texttt{uint16},
  shrinks order maintenance objects,
  which allows more of them fit per cache line,
  improving throughput.
In our implementation, we use 32-bit pointers;
  this conveniently makes the total size
  of an order maintenance object 128 bits,
  meaning that order maintenance objects
  do not split across cache lines.

The actual implementation of \texttt{OMPool} is standard;
  it stores a \texttt{pool} of memory as an \texttt{std::vector},
  in which we ensure sufficient capacity at startup.
Freed elements are placed in a separate \texttt{freed} vector,
  which is preferentially drawn from by \texttt{malloc}.
Because the objects are all the same size,
  there is no fragmentation and \texttt{malloc}/\texttt{free}
  are nearly instantaneous.
Moreover, since Spineless Traversal
  creates Order Maintenance objects in order,
  allocation patterns are extremely favorable,
  with temporally-close nodes often placed nearby in memory.

\begin{figure}
% Newlines below to align figure captions
\begin{Verbatim}[formatcom=\color{gray}, commandchars=+~!]
template<
  typename T,                   // Type of allocated object
  +color~red!~typename P=uint32_t!           +color~black!~// Integer type for "pointers"!
> struct OMPool {
  +color~red!~std::vector<T> pool;!          +color~black!~// Fast, local allocations!
  +color~black!~std::vector<P> freed;         // Rapid reuse minimizes churn!
  P malloc();                   // Straightforward
  +color~black!~void free(P p) { freed.push_back(p); }!
  +color~black!~T* addressof(P p) { return &(pool[p]); }!
}
\end{Verbatim}
\caption{A pooling allocator to reduce cache misses.
  The pointer type $P$ is smaller than a standard pointer,
    allowing order maintenance objects to be smaller
    and thereby fit tightly in a cache line.}
\label{fig:allocator}
\end{figure}

\begin{figure}
\begin{Verbatim}[formatcom=\color{gray}, commandchars={+[]}]
Label lpl = l.parent->label, rpl = r.parent->label;
Label ll = l.label, rl = r.label; uint64_t result;
asm volatile(
  "+color[black][xor   %%rbx,  %%rbx]    +color[gray][\n"]
  "+color[black][cmp   %1,     %2]       +color[gray][\n"]
  "+color[black][seta  %%bl]             +color[gray][\n"]
  "+color[black][xor   %%rax,  %%rax]    +color[gray][\n"]
  "+color[black][cmp   %3,     %4]       +color[gray][\n"]
  "+color[black][seta  %%al]             +color[gray][\n"]
  "+color[red][cmove %%rbx,  %%rax]    +color[gray][\n"]
  : "=&a"(result)
  : "r"(ll), "r"(rl), "r"(lpl), "r"(rpl));
return result;
\end{Verbatim}
\caption{
  The branchless comparison code,
    with the conditional move instruction highlighted.
  Note that, while the assembly snippet
    is seven instructions long,
    the two \texttt{xor} instructions
    are handled by the register renamer,
    and \texttt{cmp}/\texttt{seta} is fused on recent Intel/AMD CPUs.
  The assembly snippet thus executes in three cycles;
    even adding the label loads,
    comparison typically takes only five cycles.
}
\label{fig:compare}
\end{figure}

\subsection{Branchless Order Maintenance Comparison}

By using a binary heap, which has great cache locality,
priority queue pushes and pops spend basically all of their time
  comparing order maintenance objects.
Order maintenance objects are small and,
  thanks to our allocator, typically in cache.
However, order maintenance object comparison has two cases
  (same or different second-level lists)
  and the pipeline stall from the conditional
  ends up being a bottleneck.
Moreover, the branch predictor does not help much,
  because (thanks to the priority queue)
  the comparison is unpredictable.
We therefore implemented a branchless comparison function,
  relying on inline assembly \texttt{cmov} instructions,
  shown in \Cref{fig:compare};
  it executes in about five cycles of latency.
Assembly makes the implementation non-portable,
  but this function is critical for performance
  and we weren't able to make the C++ compiler
  generate comparably-fast code, despite several attempts.
If portability is a concern, we also provided a branchless version implemented using bit operations.

\subsection{Attempted, Failed Optimizations}

A number of attempted optimizations
  failed to improve performance:
\begin{itemize}
\item A hybrid of Double Dirty Bit and Spineless Traversal,
    using a summary bit for subtree dirty bit propagation
    but the priority queue for more distant jumps.
  We were unable to make switching between the two modes
    efficient enough to be competitive.
\item A splay tree instead of a min-heap.
  Slower, likely due to worse cache locality.
\item A red-black tree instead of a min-heap. Also slower.
\item A 1-based array in the min-heap, to speed up index manipulation.
  Slower.
\item 16-bit \texttt{OMPool} pointers/OM labels.
  No faster than 32 bits, and more rebalancings needed.
  We suspect that this is because 16-bit pointers/labels
  only shrink OM cells from 16 to 8 bytes,
  but load ports on modern CPUs already load 16 bytes at a time.
\item Pointer tagging to make priority queue elements smaller.
  No improvement.
\item Splitting OM \texttt{left}/\texttt{right} pointers
  from the \texttt{label} and \texttt{parent}/\texttt{children} pointers, to improve cache locality. No improvement.
\item Deallocating OM objects when the corresponding tree node is deleted, to increase cache locality. A slight slow-down.
\end{itemize}
\section{Evaluation}

We implement a fragment of web layout in Megatron
  and use it to compare Spineless Traversal
  against Double Dirty Bit
  on \NumWebsites real-world websites.

\subsection{Web Layout Fragment}
\label{sec:layout-impl}

Existing web layout implementations
  are complex and tightly coupled
  to their current invalidation strategy.
Therefore, to evaluate Spineless Traversal,
  we re-implemented web layout, 
  basing our approach on Cassius and \textsc{Medea}%
  ~\cite{cassius-1,cassius-2,yufeng-2}.
Naturally, our implementation handles only a subset
  of HTML and CSS features.
However, we took care to implement several features
  with complex invalidation behavior described below.
In total, our implementation computes approximately 50 layout fields
  over about 700 lines of Megatron DSL.

\paragraph{Box model}
Each layout node has $x$, $y$, \textsf{width} and \textsf{height} fields;
  formally, this rectangle defines its border box.
Typically a node's border box contains its children
  and doesn't overlap with siblings.
Width generally has parent-to-child dependencies,
  height generally has child-to-parent dependencies,
  while $x$ and $y$ are computed in-order.
This forms long dependency chains between elements---%
  modifying one element can eventually dirty many others---%
  but many CSS properties like
  \textsf{width}, \textsf{min-width}, and \textsf{max-width},
  and similar for \textsf{height},
  can break these dependency chains.
These properties, however, allow values like \textsf{50\%},
  which are resolved relative to the parent and thus
  still creates inter-node dependencies.
	
\paragraph{Line Breaking}
Line breaking lays out inline layout nodes (text)
  horizontally into lines.
When text reaches the right edge of its parent box,
  the next inline layout node is placed in the next line.
Line breaking thus creates control dependencies,
  where checking the parent node's width may cause layout nodes
  to move from one line to another (by changing line breaking).
Additionally, our layout algorithm
  allows different lines to have different heights
  (based on the height of the largest layout node in the line),
  which introduces a field (line height)
  that is dependent on many different nodes (each word in the line).
This also requires multiple layout passes,
  since later words in a line
  can affect the placement of earlier words
  by adjusting the text baseline.
This has a number of interesting effects for invalidation.
For example, adding a node to a line
  may or may not change the line's height,
  depending on whether the new node is tallest.
If it is, all other text on the page must move down,
  causing a lot of invalidation.

\paragraph{Display}
The \textsf{display} property changes
  whether a node acts like words (\textsf{inline})
  or paragraphs (\textsf{block}).
It can also be \textsf{none},
  in which case the layout nodes are not shown on the page
  and have almost no effect on layout;
  changing \textsf{display} between \textsf{block} and \textsf{none}
  is a common way to implement drop-down menus, pop-ups, and tool-tips.
Importantly, \texttt{<script>} and \texttt{<style>} tags
  have \textsf{display: none};
  inserting them into the page must be fast.

\paragraph{Position}
An element with \textsf{position: absolute}
  is manually assigned its $x$ and $y$ position by the web developer;
  this property is used for popups, tool-tips, and other hover effects.
It is also common
  to change the manually-assigned $x$ and $y$ positions
  from JavaScript, such as to move a tool-tip away from the cursor.
Layout nodes with absolute positioning do not affect
  the position of sibling layout nodes,
  and handling changes to $x$ and $y$ positions quickly is essential.

\paragraph{Intrinsic sizes}
Layout nodes have an ``intrinsic'' size---%
  its size without any or with all line breaks, basically---%
  which is used for, for example, absolutely-positioned elements.
Importantly, intrinsic widths are computed bottom-up,
  but are then used in the top-down width computation,
  which then affects the bottom-up height computation.
This means intrinsic sizes require the use of
  multiple layout phases.

\paragraph{Flexbox}
Flexbox layout is the most complex feature we implemented.
In flexbox layout there is flex container element
  whose children are flex items.
The width/height of flex items depends on
  the intrinsic sizes of the other flex items and
  the actual size of the flex container.
Properties like \textsf{flex-grow} and \textsf{flex-shrink}
  determine how the intrinsic sizes of the flex items
  are adjusted to match available space in the flex container.
The \textsf{max-} and \textsf{min-width}/\textsf{height} properties
  can also cap the growth/shrinkage of individual flex items.
In all, our implementation of flexbox layout
  uses 9 intermediate fields
  and requires 2 passes to compute all of them.
Note that the full web layout specification
  includes layout modes like grid layout,
  which require even more passes.

\paragraph{Miscellaneous}
We also implemented a variety of miscellaneous features,
  including automatic sizing of images and video,
  manual line breaks with the \texttt{<br>} element,
  and hidden elements like \texttt{<noscript>}
  (which are only rendered if JavaScript is disabled).
We also had to add a special case for \texttt{<svg>} elements,
  whose children describe drawing commands
  that do not participate in layout.
Finally, we also implemented the \textsf{width} and \textsf{height}
  HTML attributes
  (which behave slightly differently from the CSS properties).

\subsection{Benchmark Web Pages}

To capture web layout traces,
  we modified the Ladybird web browser
  to dump the layout tree at every rendered frame,
  including attributes and properties for each node;
  we then use a separate program to ``diff'' successive frames,
  outputting a list of insertions, deletions,
  and attribute/property changes for that frame.
The driver program then reads each frame from the trace,
  performs each modification in the frame,
  and finally invokes incremental layout for that frame.
In total, we captured traces from \NumWebsites websites,
  and those traces contain  \NumFrames frames in total;
  in our evaluation, each frame is one data point.
Note that this large number of frames,
  covering gigabytes of layout tree data,
  nonetheless represents only a few minutes of web browsing activity.
All experiments are run on a machine with
  an Intel i7-8700K CPU (8th generation)
  clocked at the standard 3.70\,GHz
  with 64\,KB L1 cache, 256\,KB L2 cache (both per core),
  and 12\,MB L3 cache (shared), plus
  32\,GB of DDR4 memory across 4 DIMMs at 3000 MT/s.

The \NumWebsites real-world websites include
  Amazon, Wikipedia, Github, Google,
  as well as a number of other
  large web pages and complex web applications
  drawn from the Alexa ranking of top websites.
A number of the authors' personal favorites are also included,
  such as Github and Lichess.

We focus on latency-sensitive interactions
  like hovering, typing, dragging, and animations.
These interactions typically
  do not require loading data over the network
  and invalidation time is thus a big determinant of their latency.
Even though the interactions may seem minor,
  it is important to note that the browser is nonetheless
  performing a significant amount of work to render them.
For example, on Wikipedia, hovering over a link
  fades a ``preview'' window in and out,
  and Wikipedia code must track and respond to mouse movements
  to hide and show the preview window at the correct time.
Moreover, there is a short, nearly-imperceptible animation
  by which the preview window slides and fades in and out of view.
Similarly, on the Lichess web page,
  our trace captures one of the authors
  stepping through a chess opening using the website's
  chess commentary tools.
The Lichess website renders the chess board using HTML elements
  and each move animates visual aids like arrows.
Text editing, an especially latency-sensitive interaction,
  was also tested.
For example, on the Google website we tested
  typing a search term letter by letter,
  with the Google website changing autocomplete suggestions
  as we typed.
Executing these interactions at low latency
  is critical for avoiding ``jank''.

\subsection{Results}

\begin{figure}
\includegraphics[scale=0.4]{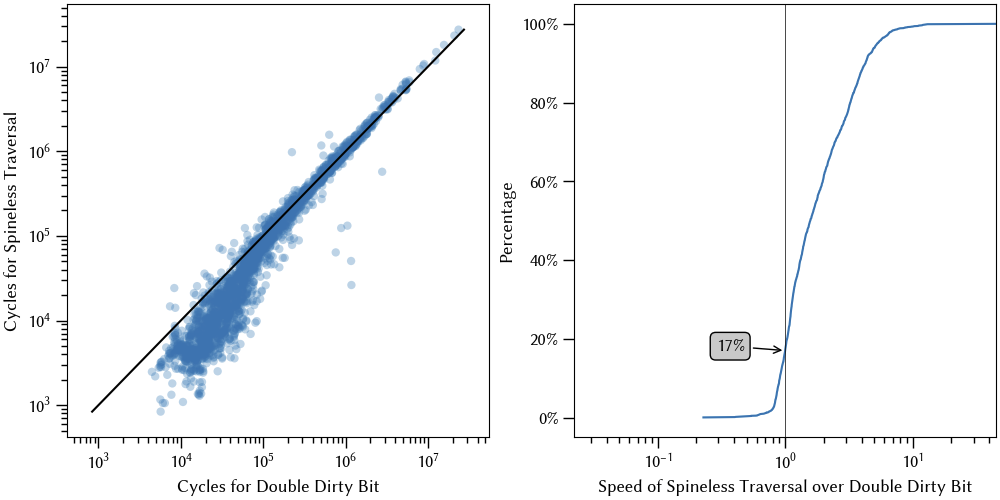}
\begin{minipage}[t]{0.48\linewidth}
\caption{
Re-layout time for all \NumFrames frames,
    with Double Dirty Bit time on the $x$ axis
    and Spineless Traversal time on $y$ axis.
The diagonal $x = y$ line shows equal time;
    points below the line are faster with Spineless Traversal
    while points above the line are faster with Double Dirty Bit.
Both axes are in log scale, meaning Spineless Traversal is often
    many faster than Double Dirty Bit.
}
\label{fig:xy}
\end{minipage}\hfill%
\begin{minipage}[t]{0.48\linewidth}
\caption{
  A CDF of the ratio between
    Double Dirty Bit time and Spineless Traversal time
    for each frame.
  The vertical line, at $10^0 = 1\times$,
    marks where both invalidation algorithms take equal time.
  To the left of the line,
    \PctSlower of frames are slower with Spineless Traversal.
  To the right of the line,
    \PctFaster of frames are faster with Spineless Traversal.
  The geometric mean speedup with Spineless Traversal
    is \MeanSpeedup.}
\label{fig:cdf}
\end{minipage}
\end{figure}

\iffalse
\begin{figure}
\begin{subfigure}{0.5\linewidth}
    \includegraphics[width=\linewidth]{DBPQLargeOverhead.png}
\end{subfigure}\hfill%
\begin{subfigure}{0.5\linewidth}
    \includegraphics[width=\linewidth]{DBPQLargeCDF.png}
\end{subfigure}
\caption{
The overhead scatter plot and the speedup cdf,
  restricted to frames where more than 1\% of fields
  are recomputed. These points typically represent loading of new contents.}
\label{fig:dbpq-large}
\end{figure}
\fi

\begin{figure}
\begin{minipage}{.59\linewidth}%
\includegraphics[width=\linewidth]{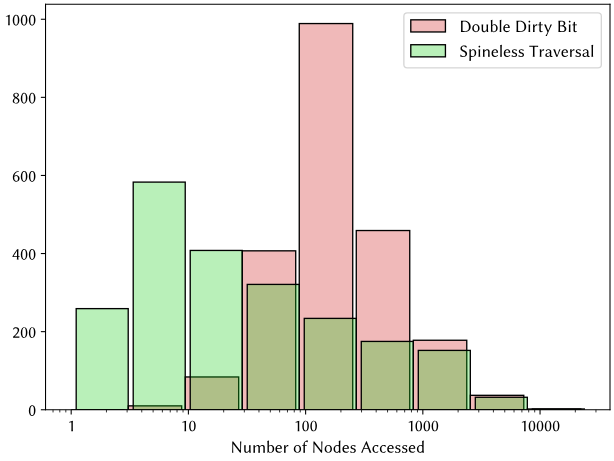}%
\caption{Histograms of Number of Nodes Accessed by Double Dirty Bit and Spineless Traversal. Double Dirty Bit access much more nodes compare to Spineless Traversal, so the latter cause much fewer cache misses.}
\label{fig:nodes-accessed}
\end{minipage}\hfill%
\begin{minipage}{.39\linewidth}%
\includegraphics[width=\linewidth]{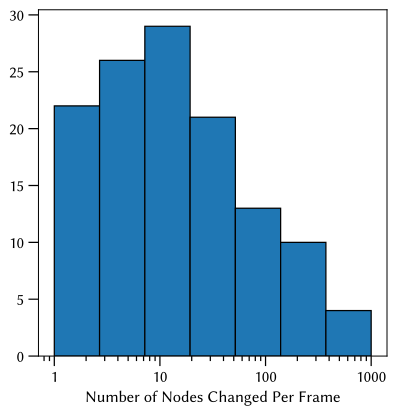}
\caption{The numbers of Twitter nodes changed externally for each frame. Most frames modify very few nodes, but a few frames insert/remove large subtrees of up to 787 nodes.}
\label{fig:case-study}
\end{minipage}
\end{figure}

\Cref{fig:xy}
  shows our results.
In it, points below the diagonal line
  are frames that are faster with Spineless Traversal,
  while points above the diagonal line
  are frames faster with Double Dirty Bit.
Most frames are below the line:
  only a few deeply-nested nodes are dirtied,
  but Double Dirty Bit makes a huge number of auxiliary accesses,
  which Spineless Traversal avoids.
By contrast, while some points are above the line,
  meaning they are slower with Spineless Traversal,
  the slowdowns are typically much less severe.
The geometric mean is a \MeanSpeedup speedup
  from Spineless Traversal,
  with only \PctSlower of frames rendered slower,
  as shown in \Cref{fig:cdf}.
\Cref{fig:nodes-accessed} shows the reason for this speedup:
  Spineless Traversal simply accesses far fewer nodes
  than Double Dirty Bit.

\Cref{fig:xy,fig:cdf} includes both
  ``overhead'' and ``evaluation'' time.
Evaluation time, as expected, is nearly identical
  between Spinless Traversal and Double Dirty Bit.
Considering overhead alone,
  Spineless Traversal is, for some frames,
  as much as $100\times$ faster than Double Dirty Bit.
For Spineless Traversal,
  overhead time was roughly one third of total runtime,
  while evaluation time was roughly two thirds,
  showing that invalidation overhead is still
  a significant determinant of latency.
Naturally, the slower Double Dirty Bit algorithm
  spends even more time in invalidation overhead.
Breaking overhead time down further for Spineless Traversal
  shows that both the priority queue
  and the order maintenance structure
  contribute to overhead, with different algorithms
  dominating for different benchmarks

A careful inspection of \Cref{fig:xy} shows
  several additional features.
The slowest frames all feature slowdowns.
This is expected: the slowest frames likely represent
  the initial page load or other ``loading'' frames,
  which we necessarily capture in our traces.
While speedups are always better than slow-downs,
  these loading frames likely follow network latency,
  so invalidation time for these frames is less important.
Meanwhile, frames where fewer nodes are invalidated
  are typically those triggered in response to
  an animation or user interaction,
  where latency is most noticeable as ``jank''.
When we restrict the data to only include frames
  where fewer than 1\% of fields are recomputed---%
  the intention being to ignore ``loading'' frames---%
  the geometric mean speedup is larger,
  at \MeanSpeedupSmall,
  and a smaller fraction of frames (\PctSlowerSmall) suffer slowdowns.
Spineless Traversal is also faster outside this subset,
  likely because the 1\% threshold is an imprecise heuristic,
  but the larger speedup 
  in this latency-critical subset is indicative.

\subsection{Case Study: Twitter}

We now focus specifically
  on our trace of Twitter (now X), a social media platform.
This trace of 125 frames captures the user
  opening the Twitter news feed,
  loading the default number of tweets,
  and scrolling down repeatedly to load more tweets.
Twitter is a large web page,
  and the tree grows to 3\thinspace700 DOM nodes
  with a depth of 53 and a fanout of 128. 
Considering all the frames in aggregate,
  Twitter sees a geometric mean speedup of $1.99\times$ 
  over the Double Dirty Bit algorithm.

Most of the 125 incremental layouts are small,
  dirtying no more than 20 nodes (Figure~\ref{fig:case-study}).
For these frames,
  Double Dirty Bit spends most of its time
  accessing auxiliary nodes.
However, the largest incremental layout
  dirties several hundred nodes.
We now discuss several common kinds of frames
  in the Twitter trace.

\paragraph{Linked Files}
Many layouts are triggered when linked files---%
  JavaScript, CSS, images, and videos---%
  finish loading.
Loading JavaScript might add
  new \texttt{<script>} and \texttt{<style>} elements to the page,
  while loading CSS files can change the CSS properties
  of existing elements.
Loading images and videos, meanwhile,
  changes intrinsic widths and heights
  from 0 to the actual image/video width/height.
Typically only one or a few nodes are dirtied,
  but these nodes are often located deep in the layout tree
  or have many siblings,
  so they have many auxiliary nodes.
Spineless Traversal thus reduces latency for these frames
  by up to $10\times$.
 
\paragraph{Lazy Loading}
Twitter uses a lazy-loading technique
  which first loads a ``shell'' page
  and then gradually adds more and more elements to the shell
  as more content is loaded over the network.
For example, the header bar, side bar, ads, and tweets
  all load separately and require separate incremental layouts.
Scrolling causes yet more content (tweets and ads) to load.
Each of these frames typically involve inserting
  a single large subtree.
Allocating the new nodes' OM objects
  (and possibly rebalancing the OM data structure)
  makes these frames difficult for Spineless Traversal
  despite its bulk insertion optimizations,
  Spineless Traversal is slower than Double Dirty Bit,
  typically by about $2\times$.
That said, the latency
  is partially hidden by the network latency
  of loading the content in the first place,
  so the slowdown here may be less critical
  than for other frames.

\paragraph{Removal}
The Twitter application also occasionally removes
  subtrees that are no longer visible to the user,
  like offscreen content.
Spineless Traversal handles these removals
  much faster than Double Dirty Bit,
  often by $5\times$ or more,
  precisely because these removals
  do not affect what the user sees on the screen.
Twitter also sometimes removes
  individual \texttt{<script>} and \texttt{<style>} elements
  that don't affect the page;
  here Spineless Traversal's speedup
  is smaller, approximately $2\times$,
  as multiple elements are removed at once,
  amortizing the auxiliary accesses in Double Dirty Bit. 

Moreover,
  some frames mix file loading, lazy loading, and removals,
  probably at the whim of the task scheduler.
Often many images load in at once.
In this case the time taken for Spineless Traversal 
  basically sums over the time for each individual change,
  while Double Dirty Bit can amortize the cost
  of traversing auxiliary nodes.
These frames are often small,
  so Spineless Traversal's speed-ups are still substantial,
  but probably smaller than if each modification
  was laid out in its own frame.

\section{Related Work}

\paragraph{Incremental Computation}
Speeding up computations
  by reusing previously-computed results
  is a long-studied topic in computer science broadly~\cite{memo}
  and programming language theory in particular;
  \citet{IC-Survey} and \citet{IC-bib}
  give thorough surveys of the field.
The recent Self-Adjusting Computation (SAC)~\cite{SAC} framework
  proposes incrementalizing arbitrary computations,
  including a cost semantics~\cite{SACCost},
  optimizations for data structure operations~\cite{SACTrace},
  and opportunities for parallelization~\cite{PSAC}.
The Adapton framework~\cite{Adapton}
  aims at \emph{demand-driven} incremental computation,
  and allows manually-specified annotations~\cite{AdaptonName}
  for greater reuse.
While this prior work focuses on general-purpose computations,
  Spineless Traversal is focused on a particularly critical application: web browser layout.
This application-specific focus has precedent:
  \citet{ICC} speed up memoization for functional programs over lists
  using ``chunky decomposition'',
  while differential dataflow~\cite{DDF},
  which incrementalizes relational algebra in databases,
  is prominent in industry.

\citet{TR1} wrote the earliest work
  on incremental evaluation of attribute grammars,
  motivated by syntax-directed editors;
  later work~\cite{TR2} allows references
  to non-neighbor attributes.
However, these early papers require recomputation
  immediately after every tree change,
  whereas web browsers, our target application,
  batch multiple updates
  and perform layout only once per frame.
The standard in web browsers is instead
  the Double Dirty Bit algorithm,
  described in industry publications~\cite{tali-garseil}
  and textbooks~\cite{wbe}.
  
The formal methods community has put significant effort
  into formalizing web page layout.
\citet{meyerovich-1} proposed using attribute grammars,
  similar to the DSL in \Cref{fig:dsl},
  for formalizing web-like layout rules.
Later work~\cite{yufeng-1} proposes
  synthesizing schedules from the attribute grammar rules,
  including proposals~\cite{meyerovich-2,meyerovich-3}
  to use parallel schedules to further improve layout performance,
  though these proposals have not proven practical~\cite{servo-no-parallel}.
The Cassius project~\cite{cassius-1}
  formalizes a significant fragment of CSS~2.1
  using an attribute-grammar-like formalism.
Our layout implementation is based on Cassius.
Later work also proposed
  using the Cassius formalism to verify web page layouts~\cite{cassius-2},
  including in a custom proof assistant~\cite{cassius-3}.
However, none of these works investigate incremental layout.
By contrast, the \textsc{Medea} project~\cite{yufeng-2}
  proposed synthesizing incremental layout algorithms
  by automatically synthesizing dirty bit propagation code.
Our work extends \textsc{Medea} by exploring
  optimized incremental traversal algorithms.

\begin{acks}
We thank Andreas Kling and Chris Harrelson for many conversations
discussing implementation challenges in layout invalidation and Ian
Kilpatrick for his discussion of reflow roots. Ryan
Stutsman and Anton Burtsev for helped with measuring low-level performance.
We also thank the anonymous reviewers for their many comments and
suggestions, which greatly improved the paper. This work was funded by
the National Science Foundation under Grant No. CCF-2340192 and by a
Google Faculty Research Award.
\end{acks}
\bibliographystyle{ACM-Reference-Format}
\bibliography{main}
\end{document}